\documentclass[fleqn,twoside]{article}
\usepackage{espcrc2}

\usepackage{graphicx}
\usepackage[figuresright]{rotating}

\newcommand{\AmS}{{\protect\the\textfont2
  A\kern-.1667em\lower.5ex\hbox{M}\kern-.125emS}}

\title{Chiral Symmetry Restoration and Deconfinement to Quark Matter in Neutron Stars }

\author{V.A. Dexheimer\address{Physics Department, Gettysburg College , Gettysburg, United States}%
        \thanks{vantoche@gettysburg.edu},
        S. Schramm\address{CSC, FIAS, ITP, Johann Wolfgang Goethe University, Frankfurt am Main, Germany}%
        \thanks{schramm@th.physik.uni-frankfurt.de}}
       
\begin{document}

\begin{abstract}
We describe an extension of the hadronic SU(3) non-linear sigma model to include quarks. As a result, we obtain an effective model which interpolates between hadronic and quark degrees of freedom. The new parameters and the potential for the Polyakov loop (used as the order parameter for deconfinement) are calibrated in order to fit lattice QCD data and reproduce the QCD phase diagram. Finally, the equation of state provided by the model, combined with gravity through the inclusion of general relativity, is used to make predictions for neutron stars. 
\vspace{1pc}
\end{abstract}

\maketitle

\section{Introduction}
Dense matter hadronic models, such as the ones used to describe neutron stars, work in a determined range of energies \cite{Glendenning:1991ic,Weber:1989uq,Schaffner:1995th}. They can, in addition, fulfill some symmetries of the underlying theory (QCD) such as chiral symmetry \cite{chiral2,Heide:1993yz,Carter:1995zi} and scale invariance \cite{Bonanno:2008tt,Bonanno:2009fg} but they do not include deconfinement to quark matter. On the other hand, the broadly used bag-model \cite{bag0} includes quark degrees of freedom but does not include chiral symmetry. Models such as the quark-NJL and quark sigma-models \cite{Nambu:1961tp,Nambu:1961fr,Bub} include that symmetry, but do not directly incorporate hadronic degrees of freedom or quark confinement.

Our goal is to construct an effective model that contains hadronic and quark degrees of freedom present at different densities/temperatures but that can also coexist in a mixed phase. This allows the deconfinement phase transition to be a steep first order as well as a smooth crossover and cases in between. The last two possibilities, despite being predicted by lattice QCD, cannot be achieved by the usual procedure that connects, by hand, two different models with separate equations of state for hadronic and quark matter at the chemical potential at which the pressure of the quark EOS exceeds the hadronic one \cite{hybrid1}.

In order to achieve this goal, we employ a single model for the hadronic and quark phases. Our extension of the  hadronic SU(3) non-linear sigma model also includes quark degrees of freedom in a spirit similar to the PNJL model \cite{PNJL}, in the sense that it is a non-linear sigma model that uses the Polyakov loop $\Phi$ as the order parameter for the deconfinement. This is a quite natural idea, since the Polyakov loop is related to the Z(3) symmetry, that is spontaneously broken by the presence of quarks. In QCD the Polyakov loop $\Phi$ is defined via $\Phi=\frac13$Tr$[\exp{(i\int d\tau A_4)}]$, where $A_4=iA_0$ is the temporal component of the SU(3) gauge field.

\section{The Model}
The Lagrangian density of our non-linear sigma model becomes:
\begin{eqnarray}
&L = L_{Kin}+L_{Int}+L_{Self}+L_{SB}-U,&
\end{eqnarray}
where besides the kinetic energy term for hadrons, quarks, and leptons (included to insure charge neutrality)
the terms:
\begin{eqnarray}
&L_{Int}=-\sum_i \bar{\psi_i}[\gamma_0(g_{i\omega}\omega+g_{i\phi}\phi+g_{i\rho}\tau_3\rho)\nonumber&\\&+m_i^*]\psi_i,&
\end{eqnarray}
\newline

\begin{eqnarray}
&L_{Self}=-\frac{1}{2}(m_\omega^2\omega^2+m_\rho^2\rho^2+m_\phi^2\phi^2)\nonumber&\\&
-g_4\left(\omega^4+\frac{\phi^4}{4}+3\omega^2\phi^2+\frac{4\omega^3\phi}{\sqrt{2}}+\frac{2\omega\phi^3}{\sqrt{2}}\right)\nonumber&\\&
+\frac{1}{2}k_0(\sigma^2+\zeta^2+\delta^2)-k_1(\sigma^2+\zeta^2+\delta^2)^2\nonumber&\\&-k_2\left(\frac{\sigma^4}{2}+\frac{\delta^4}{2}
+3\sigma^2\delta^2+\zeta^4\right)\nonumber&\\&
-k_3(\sigma^2-\delta^2)\zeta-k_4\ \ \ln{\frac{(\sigma^2-\delta^2)\zeta}{\sigma_0^2\zeta_0}},&
\end{eqnarray}
\begin{eqnarray}
&L_{SB}= m_\pi^2 f_\pi\sigma+\left(\sqrt{2}m_k^ 2f_k-\frac{1}{\sqrt{2}}m_\pi^ 2 f_\pi\right)\zeta,&
\end{eqnarray}

\noindent represent (in mean field approximation) the interactions between baryons (and quarks) and vector and scalar mesons, the self-interactions of the scalar and vector mesons and an explicit chiral symmetry breaking term, responsible for producing the masses of the pseudo-scalar mesons. The Polyakov-loop potential $U$ will be discussed in detail below. The underlying flavor symmetry of the model is SU(3) and the index $i$ denotes the baryon octet and the three light quarks. The mesons included are the vector-isoscalars $\omega$ and $\phi$, the vector-isovector $\rho$, the scalar-isoscalars $\sigma$ and $\zeta$ (strange quark-antiquark state) and  the scalar-isovector $\delta$. The isovector mesons affect isospin-asymmetric matter and are consequently important for neutron star physics. The mesons are treated as classical fields within the mean-field approximation \cite{MFT}. A detailed discussion of the purely hadronic part of the Lagrangian,  that describes very well nuclear saturation properties as well as nuclei properties can be found in \cite{chiral1,chiral2,eu}.

Finite-temperature calculations include a heat bath of hadronic and quark quasiparticles within the grand canonical potential of the system. Finite temperature calculations also include a gas of free pions and kaons. As they have very low mass, they dominate the low density/ high temperature regime. All calculations were performed considering zero net strangeness except the zero temperature star matter case. The reason for such exception is that the time scale in neutron stars is large enough for strangeness not to be conserved.

The effective masses of the baryons and quarks are generated by the scalar mesons except for a small explicit mass term $M_0$ (equal to $150$ MeV for nucleons, $354$ MeV for hyperons, $5$ MeV for up and down quarks and $150$ MeV for strange quarks) and the term containing the Polyakov field $\Phi$:
\begin{eqnarray}
&M_{B}^*=g_{B\sigma}\sigma+g_{B\delta}\tau_3\delta+g_{B\zeta}\zeta\nonumber&\\&+M_{0_B}+g_{B\Phi} \Phi^2,&
\label{5}
\end{eqnarray}
\begin{eqnarray}
&M_{q}^*= g_{q\sigma}\sigma+g_{q\delta}\tau_3\delta+g_{q\zeta}\zeta\nonumber&\\&+M_{0_q}
+g_{q\Phi}(1-\Phi).&
\label{6}
\end{eqnarray}
With the increase of temperature/density, the $\sigma$ field (non-strange chiral condensate) decreases its value, causing the effective masses of the hadrons to decrease towards chiral symmetry restoration. The Polyakov loop assumes non-zero values with the increase of temperature/density and, due to its presence in the baryons effective mass (Eq.~\ref{5}), suppresses their propagation. On the other hand, the presence of the Polyakov field in the effective mass of the quarks, included with a negative sign (Eq.~\ref{6}), insures that they will not be present at low temperatures/densities.

\begin{figure}[htb]
\includegraphics[clip,trim=0  0 0 2,width=7.2cm]{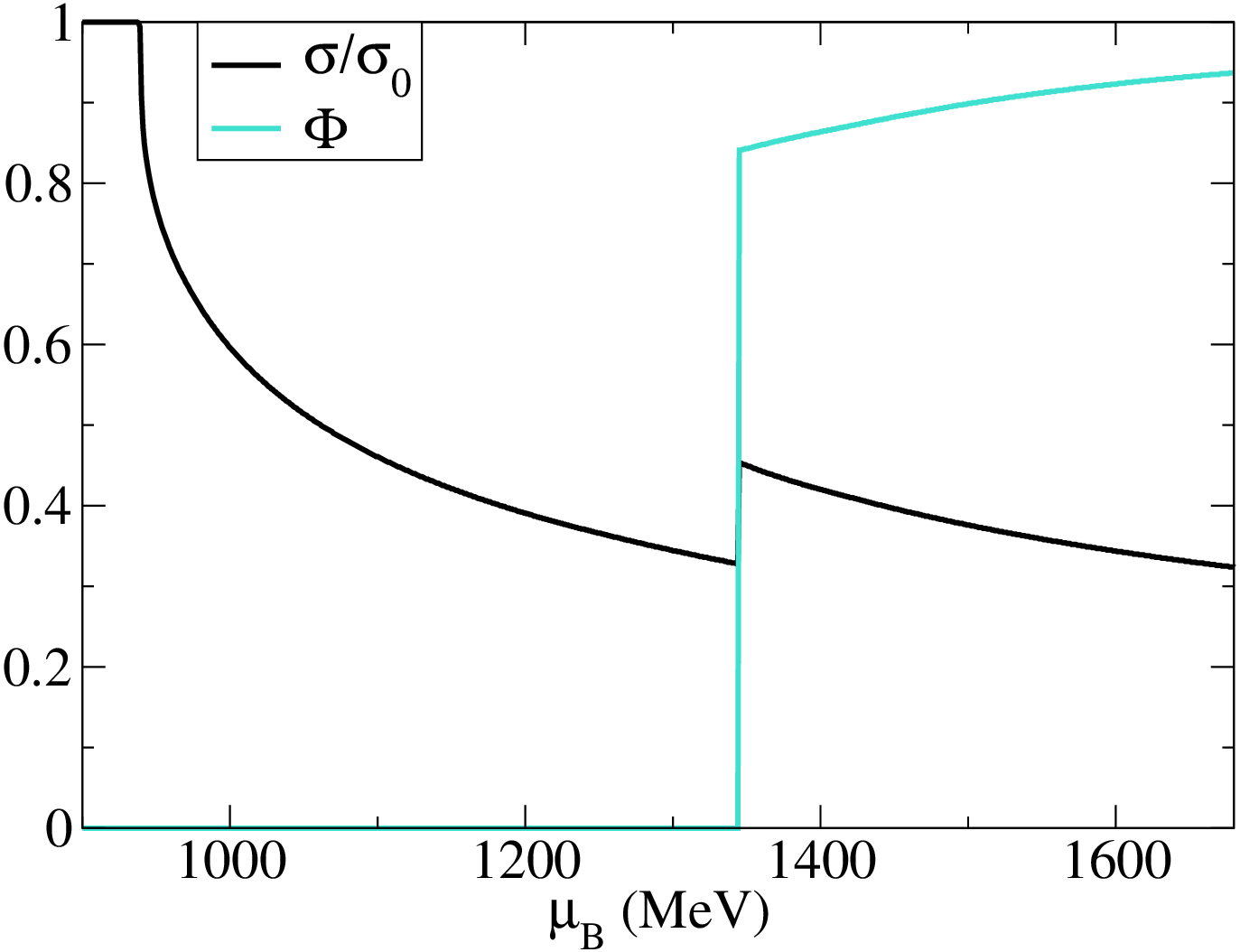}
\caption{\label{Pol}Order parameters for chiral restoration and deconfinement
to quark matter for star matter at zero temperature.}
\end{figure}

The behavior of the order parameters of the model is shown in Fig.~\ref{Pol} for neutron star matter at zero temperature. The calculations for neutron stars assume charge neutrality (which is essential for their stability) and beta equilibrium. In this case, the chiral symmetry restoration, which is a crossover for purely hadronic matter, turns into a first order phase transition by the influence of the strong first-order transition to deconfined matter. The fact that the value of the chiral condensate increases at a certain chemical potential is due to the fact that it is connected to the baryon density value that decreases at the phase transition (the baryonic number of quarks is one third of the hadronic one). The model is consistent in the sense that both order parameters are related.

\begin{figure}[htb]
\centering
\includegraphics[clip,trim=0  0 0 1,width=7.9cm]{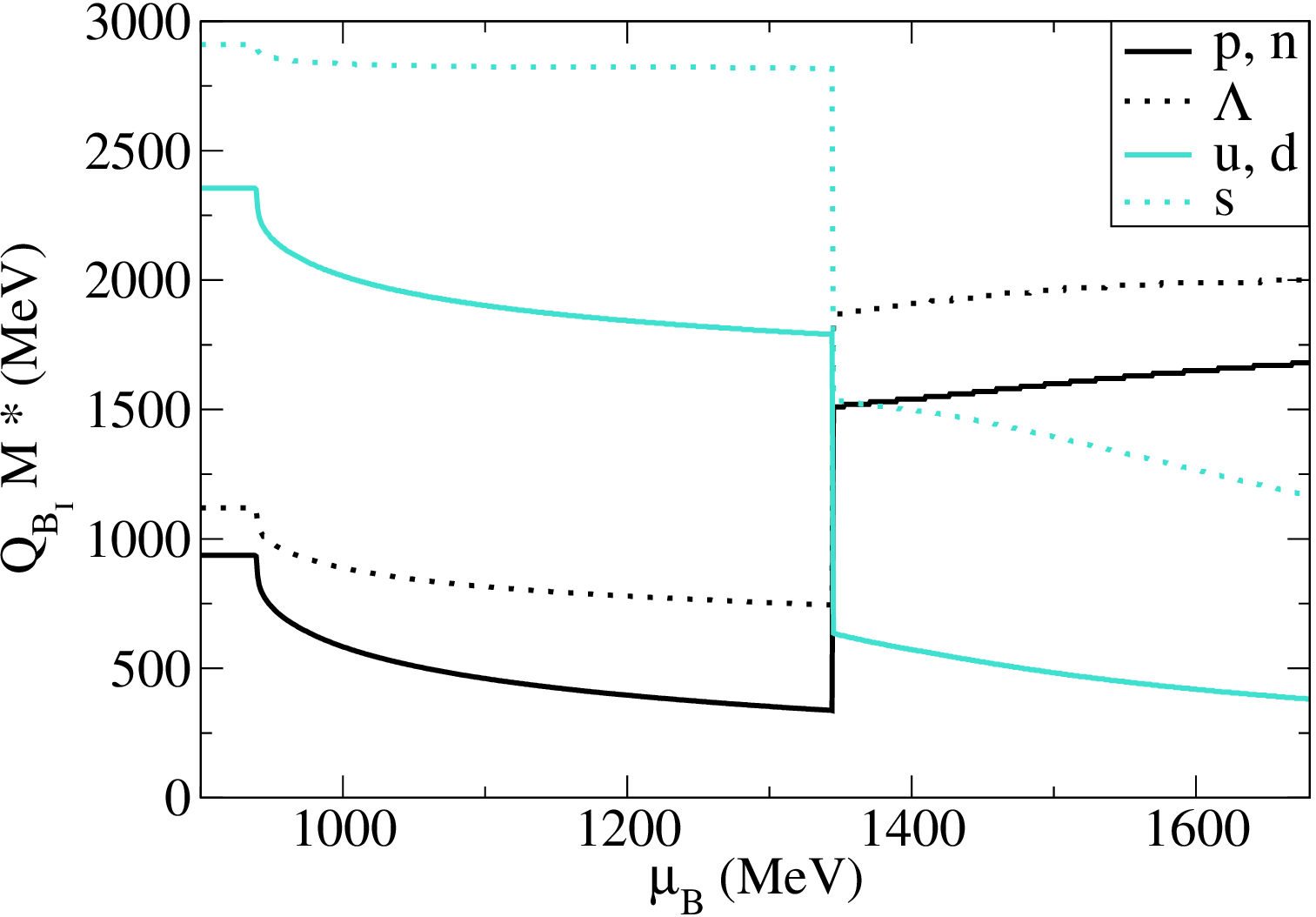}
\caption{\label{Meff}Effective normalized mass of baryons and quarks for star matter at zero temperature.}
\end{figure}

The coupling constants related to the Polyakov loop in the effective mass formulas (Eq.~\ref{5} and Eq.~\ref{6}) are high but still finite. This causes the effective masses of the degrees of freedom not effectively present in each phases to be high but also finite (Fig.~\ref{Meff}). The effective normalized masses of the particles are directly related to the onset these particles appearance in the system.

The potential $U$ for the Polyakov loop reads:
\begin{eqnarray}
&U=(a_0T^4+a_1\mu^4+a_2T^2\mu^2)\Phi^2\nonumber&\\&+a_3T_0^4\ln{(1-6\Phi^2+8\Phi^3-3\Phi^4)}.&
\end{eqnarray}
It is based on \cite{Ratti1,Ratti2} and adapted to also include terms that depend on the chemical potential. The two extra terms (that depend on the chemical potential) are not unique, but the most simple natural choice  to reproduce the main features of the phase diagram at finite densities. The coupling constants for the baryons (already shown in \cite{eu}) are chosen to reproduce the vacuum masses of the baryons and mesons, nuclear saturation properties, and asymmetry energy as well as the hyperon potentials. The vacuum expectation values of the scalar mesons are constrained by
reproducing the pion and kaon decay constants. The coupling constants for the quarks ($g_{q\omega}=0$, $g_{q\phi}=0$, $g_{q\rho}=0$, $g_{q\sigma}=-3.00$, $g_{q\delta}=0$, $g_{q\zeta}=-3.00$, $T_0=270/200$ MeV, $a_0=-1.85$, $a_1=-1.44$x$10^{-3}$, $a_2=-0.08$, $a_3=-0.40$, $g_{B\Phi}=1500$ MeV, $g_{q\Phi}=500$ MeV) are chosen to reproduce lattice data as well as known information about the phase diagram.

 \begin{figure}[htb]
\centering
\includegraphics[clip,trim=0  6 0 0,width=8.0cm]{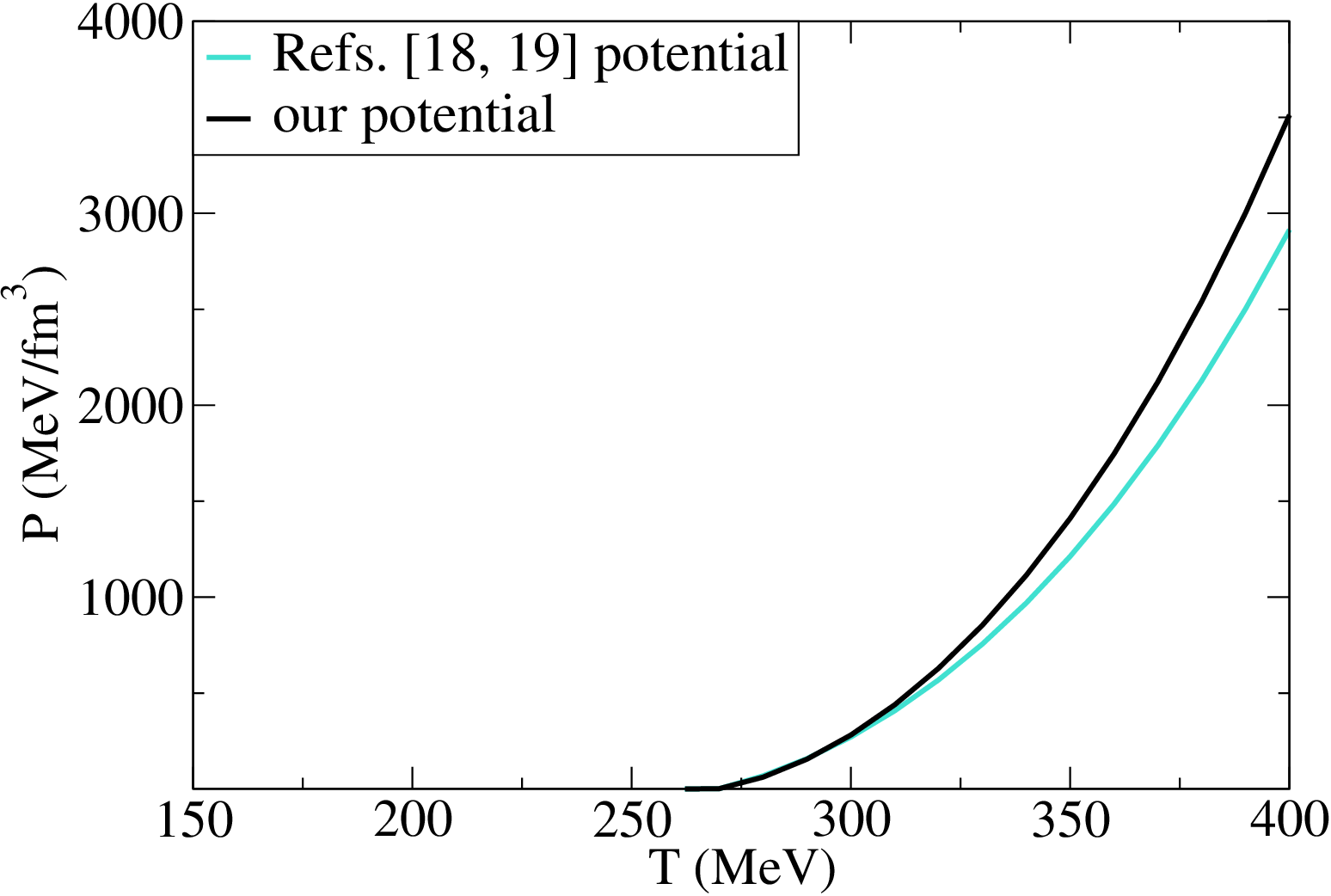}
\caption{\label{p}Pressure for the quenched case.}
\end{figure}

\begin{figure}[htb]
\centering
\includegraphics[clip,trim=0  6 0 2,width=7.2cm]{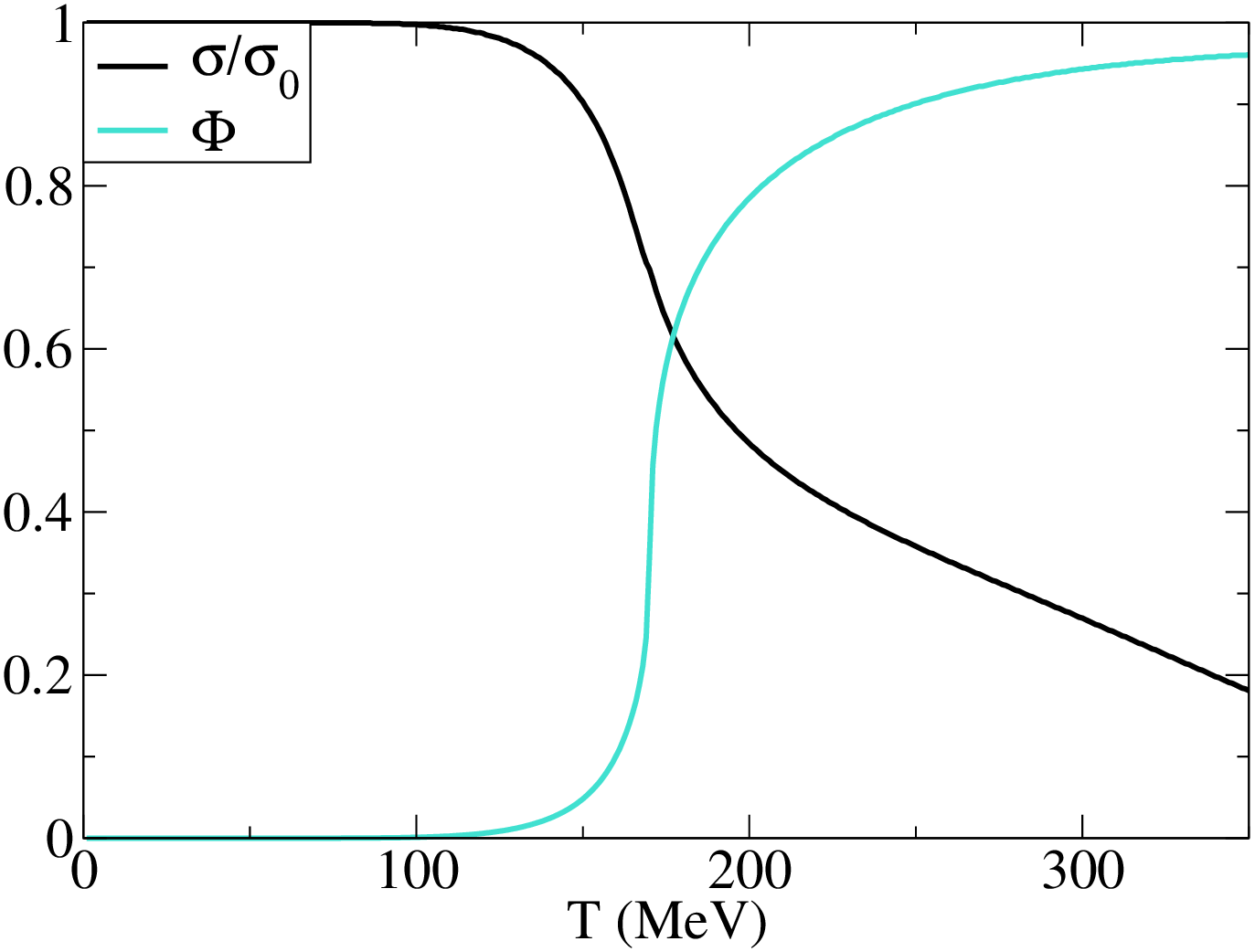}
\caption{\label{s}Order parameters for chiral symmetry restoration and deconfinement
to quark matter at zero chemical potential.}
\end{figure}

More specifically, the parameters $a_0$ and $a_3$ of the Polyakov potential were fit to reproduce the same pressure functional as the one in the model presented in Refs. \cite{Ratti1,Ratti2} for the quenched case for the range of temperature of interest (Fig.~\ref{p}). In this way we ensure the agreement with lattice at the zero chemical potential limit. The parameter $T_0$ was chosen to be $270$ MeV for quenched calculations reproducing a very strong first order phase order transition at $270$ MeV temperature (as in Ref. \cite{Ratti1}). The parameter $T_0$ was rescaled for the calculations including quarks and hadrons also as in Ref. \cite{Ratti1}. In our case it changed from $T_0=270$ to $200$ MeV reproducing, at zero chemical potential, a peak on the susceptibility for the Polyakov loop at $T=190$ MeV (Fig.~\ref{s}).

The functional form of the potential in our model is very similar to the one in Refs. \cite{Ratti1,Ratti2} with a leading term proportional to temperature to the fourth power. At the moment, the other powers of temperature (third and second) were not considered to avoid a large number of parameters. In our case, the potential also contains a term proportional to chemical potential to the fourth power, with parameter $a_2$ (chosen to reproduce a phase transition from hadronic to quark matter at a value of four times saturation density for star matter at zero temperature), and a mixed term between temperature and chemical potential, with parameter $a_1$ (chosen to reproduce a critical endpoint at chemical potential $\mu_c=354$ MeV and temperature $T_c=167$ MeV for symmetric matter and zero net strangeness in accordance with lattice data from Fodor and Katz Ref. \cite{fodor}). Together, these parameters lead the model to reproduce the QCD phase diagram at large densities. More complicated structures for the Polyakov potential will be analyzed in future work.

 \begin{figure}[htb]
\centering
\includegraphics[clip,trim=0  1 0 1,width=7.9cm]{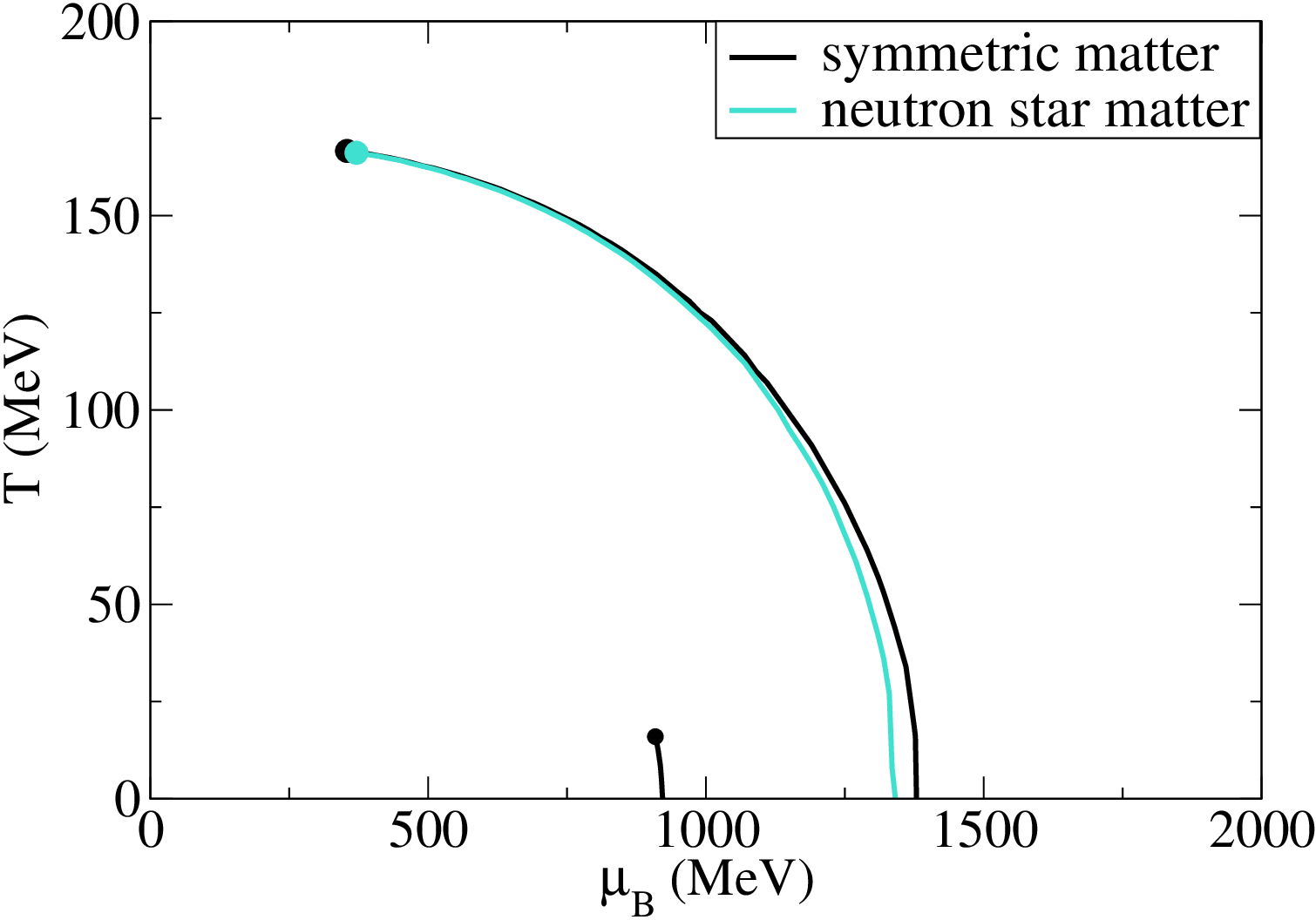}
\caption{\label{phase}Phase diagram. The lines represent first order transitions and the circles mark the critical end-points.}
\end{figure}
 
As can be seen in Fig.~\ref{phase} the transition from hadronic to quark matter obtained is a crossover for small and vanishing chemical potentials. Beyond the critical end point, a first order transition line begins. The critical temperatures for chiral symmetry restoration coincide with the ones for deconfinement. Since the model is able to reproduce nuclear matter saturation at realistic values for the saturation density, nuclear binding energy, as well as the compressibility and asymmetry energy, we also reproduce results at low densities for the nuclear matter liquid-gas phase transition.

\section{Application to Neutron Stars}

In the same way we used lattice QCD data to calibrate and test the model for high temperatures and low chemical potentials, we can calculate neutron star properties to test the model in the high-density/low-temperature regime. It is important to notice that up to this point, the charge neutrality was considered to be local, meaning that each phase was separately charge neutral. At finite temperature the two phases contain mixtures of hadrons and quarks, which are dominated by hadrons or quarks, depending on the respective phase. At vanishing temperature there is no mixture, i.e. the system exhibits a purely hadronic or purely quark phase.

It is important to notice that the quarks are totally suppressed in the hadronic phase but the hadrons are suppressed in the quark phase until a certain chemical potential (above $1700$ MeV for $T=0$).  This threshold, which is higher than the density in the center of neutron stars, establishes a limit for the applicability of the model.

\begin{figure}[htb]
\centering
\includegraphics[clip,trim=0  1 0 2,width=8.1cm]{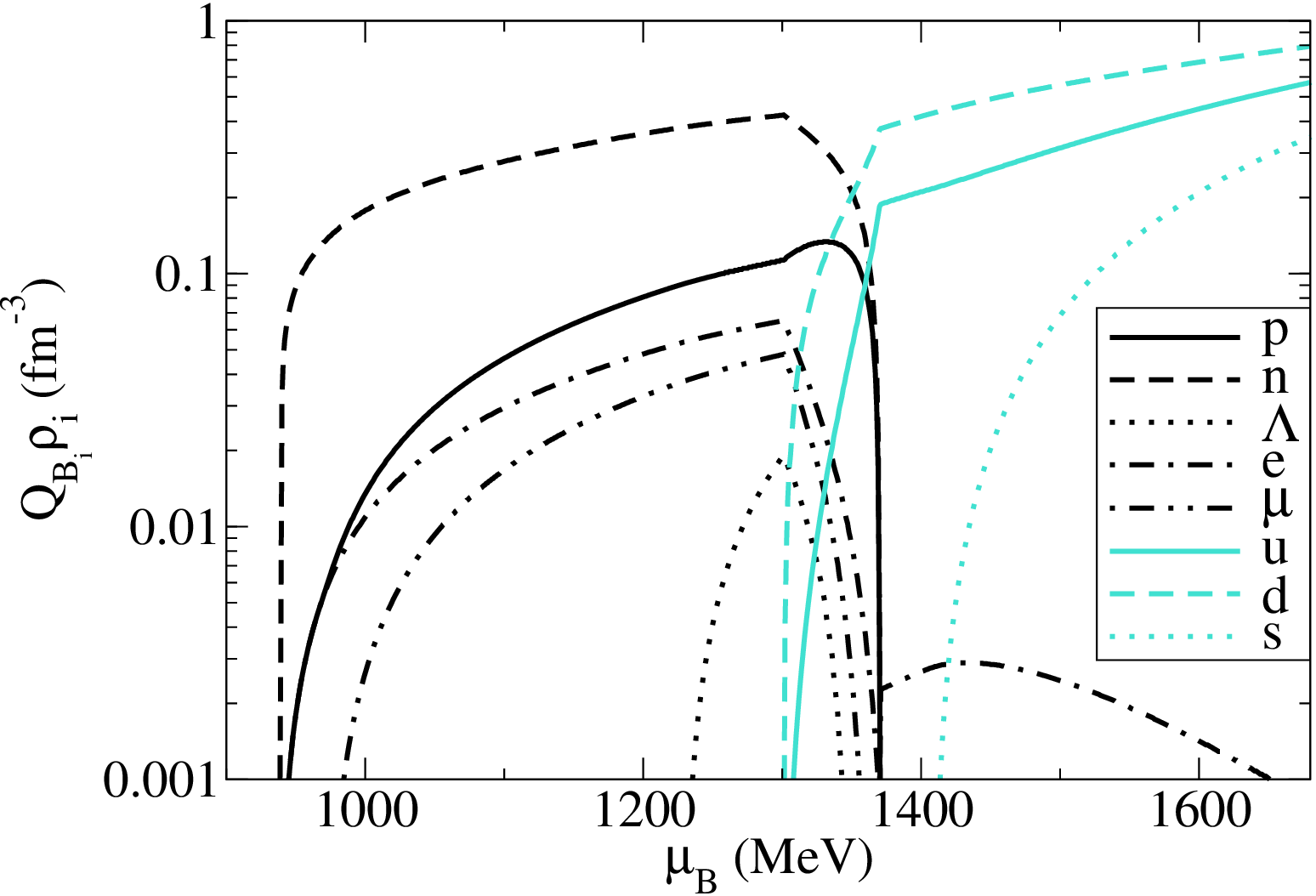}
\caption{\label{popgib}Population (baryonic densities) for star matter at zero temperature using global charge neutrality.}
\end{figure}

For a more realistic approach we allow the two phases to be charge neutral  when combined (global charge neutrality) following \cite{Glendenning:1992vb}. In this case, the particle densities change in the coexistence region causing quarks and baryons to appear and vanish in a smoother way (Fig.~\ref{popgib}).

The density of electrons and muons is significant in the hadronic phase but not in the quark phase. The reason for this behavior is that because the down and strange quarks are also negatively charged, electrons are not necessary for charge neutrality, and only a small amount of leptons remains to ensure beta equilibrium. The hyperons, in spite of being included in the calculation, are suppressed by the  appearance of the quark phase. Only a very small amount of $\Lambda$'s appear right before the phase transition. The strange quarks appear after the other quarks and also do not make substantial changes in the system.

\begin{figure}[htb]
\centering
\vspace{1.0cm}
\includegraphics[clip,trim=0  6 0 1,width=7.7cm]{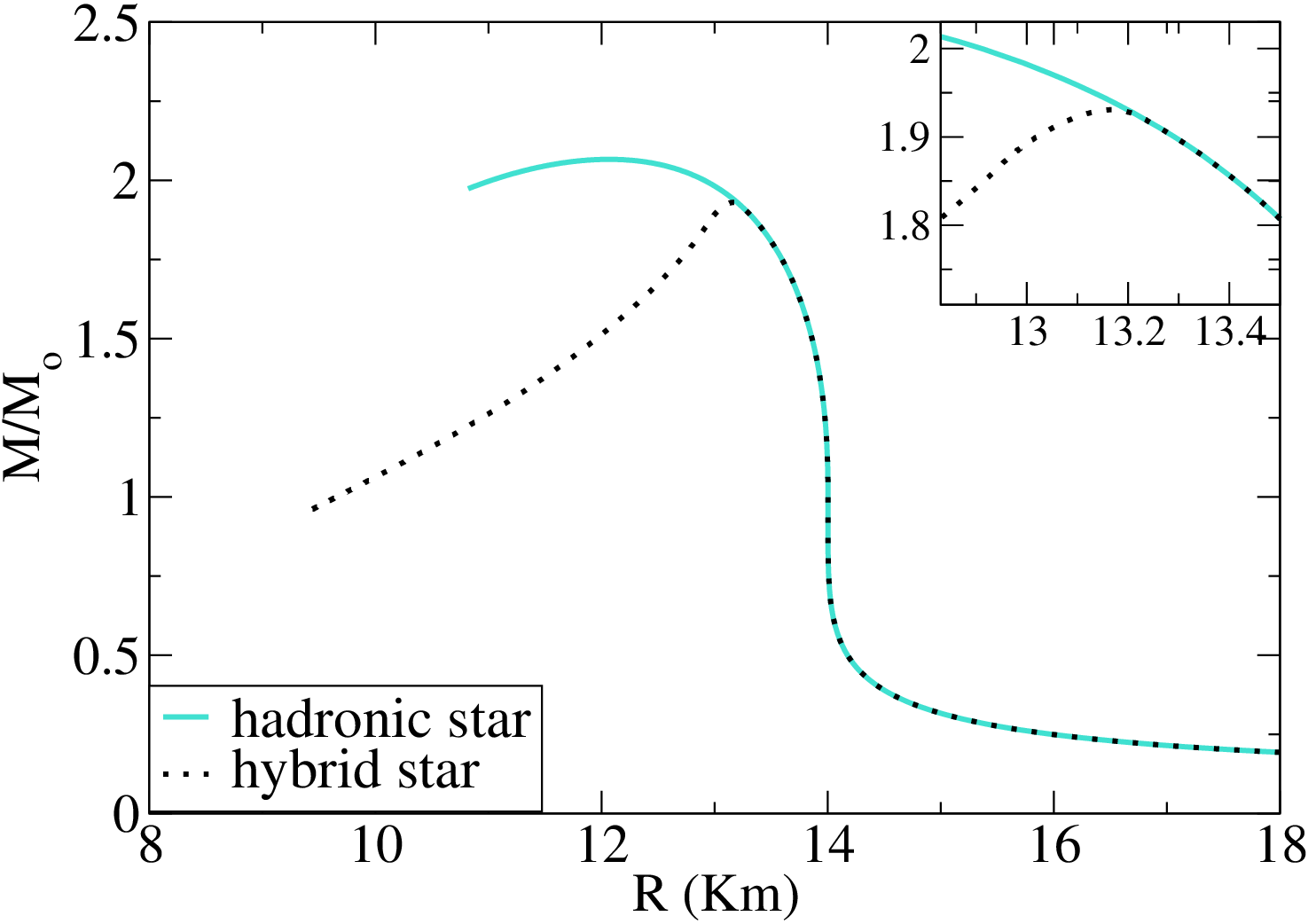}
\caption{\label{mass}Mass-radius diagram}
\end{figure}

The possible neutron star masses and radii are calculated by solving the Tolmann-Oppenheimer-Volkof equations \cite{tov1,tov2}. The solutions for hadronic (same model but without quarks) and hybrid stars are shown in Fig.~\ref{mass}, where besides our equation of state for the core, a separate equation of state was used for the crust \cite{crust}. The maximum mass supported against gravity in our model is $2.1M_\odot$ in the first case and approximately $2.0M_\odot$ in the second. The predicted radii are in the observed range, being practically the same for hadronic or hybrid stars.

Although the stars that contain a phase of pure quark matter in our model are not stable, stars that contain a core of mixed phase (surrounded by a hadronic shell) are. In this case, the mixed phase could extend up to approximately $2$ km of radius but with only a small quark fraction.

\section{Conclusions}

In spite of the fact that our model is relatively simple, it is the only one able to take into account different degrees of freedom and consequently allow steep as well as smooth transitions between different phases.  The model is in accordance with lattice QCD data and the phase diagram it reproduces is able to describe a broad variety of regimes, from compact stars to heavy ion collisions. Calculations along this line are in progress \cite{soon}. 

We conclude that our model is suitable for the description of neutron stars, since it predicts masses and radii in agreement with observations. Although it does not predict stable stars containing pure quark matter, the model allows stars that contain a core of mixed phase that can extend up to approximately $2$ km of radius. Even in this case, the reduced maximum star mass is still higher than the most massive pulsars observed. 

A major advantage of our work compared to other studies of hybrid stars is that  we can study in detail the way in which chiral symmetry is restored and the way deconfinement occurs at high temperature/density. Since the properties of the physical system, e.g. the density of particles in each phase, are directly connected to the Polyakov loop it is not surprising that we obtain different results in a combined description of the degrees of freedom compared to a simple matching of two separate equations of state.


\begin{thebibliography}{99}

\bibitem{Glendenning:1991ic}
  N.~K.~Glendenning, F.~Weber and S.~A.~Moszkowski,
  Phys.\ Rev.\  C {\bf 45}, 844 (1992).

\bibitem{Weber:1989uq}
  F.~Weber and M.~K.~Weigel,
  Nucl.\ Phys.\  A {\bf 505}, 779 (1989).

\bibitem{Schaffner:1995th}
  J.~Schaffner and I.~N.~Mishustin,
  Phys.\ Rev.\  C {\bf 53}, 1416 (1996)
  [arXiv:nucl-th/9506011].

\bibitem{chiral2}
  P.~Papazoglou, D.~Zschiesche, S.~Schramm, J.~Schaffner-Bielich, H.~Stocker and W.~Greiner,
  Phys.\ Rev.\  C {\bf 59}, 411 (1999).

\bibitem{Heide:1993yz}
  E.~K.~Heide, S.~Rudaz and P.~J.~Ellis,
  Nucl.\ Phys.\  A {\bf 571}, 713 (1994)
  [arXiv:nucl-th/9308002].

\bibitem{Carter:1995zi}
  G.~W.~Carter, P.~J.~Ellis and S.~Rudaz,
  Nucl.\ Phys.\  A {\bf 603}, 367 (1996)
  [Erratum-ibid.\  A {\bf 608}, 514 (1996)]
  [arXiv:nucl-th/9512033].

\bibitem{Bonanno:2008tt}
  L.~Bonanno and A.~Drago,
  arXiv:0805.4188 [nucl-th].
  
  \bibitem{Bonanno:2009fg}
  L.~Bonanno,
  arXiv:0909.0924 [nucl-th].

\bibitem{bag0} F. Weber, Prog. Part. and Nucl. Phys. {\bf 54}, 193 (2005), and references therein.
%

\bibitem{Nambu:1961tp}
  Y.~Nambu and G.~Jona-Lasinio,
  Phys.\ Rev.\  {\bf 122}, 345 (1961).

\bibitem{Nambu:1961fr}
  Y.~Nambu and G.~Jona-Lasinio,
  Phys.\ Rev.\  {\bf 124}, 246 (1961).

\bibitem{Bub}
  M.~Buballa,
  Phys.\ Rept.\  {\bf 407}, 205 (2005)
  [arXiv:hep-ph/0402234].


\bibitem{hybrid1}
  H.~Heiselberg, C.~J.~Pethick and E.~F.~Staubo,
  Phys.\ Rev.\ Lett.\  {\bf 70}, 1355 (1993).

\bibitem{PNJL}
  K.~Fukushima,
  Phys.\ Lett.\  B {\bf 591}, 277 (2004)
  [arXiv:hep-ph/0310121].


\bibitem{MFT} J.D. Walecka, {\sl Theoretical Nuclear And Subnuclear Physics} World Scientific Publishing Company; 2nd edition (2004).

\bibitem{chiral1}
  P.~Papazoglou, S.~Schramm, J.~Schaffner-Bielich, H.~Stocker and W.~Greiner,
  Phys.\ Rev.\  C {\bf 57}, 2576 (1998).

\bibitem{eu}
  V.~Dexheimer and S.~Schramm,
  Astrophys.\ J.\ {\bf 683}, 943 (2008).

\bibitem{Ratti1}
  C.~Ratti, M.~A.~Thaler and W.~Weise,
  Phys.\ Rev.\  D {\bf 73}, 014019 (2006)

\bibitem{Ratti2}
  S.~Roszner, C.~Ratti and W.~Weise,
  Phys.\ Rev.\  D {\bf 75}, 034007 (2007)

\bibitem{fodor}
  Z.~Fodor and S.~D.~Katz,
  JHEP {\bf 0404}, 050 (2004)
  [arXiv:hep-lat/0402006].

\bibitem{Glendenning:1992vb}
  N.~K.~Glendenning,
  Phys.\ Rev.\  D {\bf 46}, 1274 (1992).

\bibitem{tov1}
  R.~C.~Tolman,
  Phys.\ Rev.\  {\bf 55}, 364 (1939).

\bibitem{tov2}
  J.~R.~Oppenheimer and G.~M.~Volkoff,
  Phys.\ Rev.\  {\bf 55}, 374 (1939).

\bibitem{crust}
 G.~Baym, C.~Pethick and P.~Sutherland,
  Astrophys.\ J.\  {\bf 170}, 299 (1971).

\bibitem{soon}
J.~Steinheimer, V.~Dexheimer, H.~Petersen, M.~Bleicher, S.~Schramm and H.~Stoecker,
  arXiv:0905.3099 [hep-ph].



\end{thebibliography}
\end{document}